\documentclass[a4paper,12pt]{article}
\usepackage{amsmath,amsfonts,amssymb,cite}
\usepackage{graphicx
}

\begin{document}

\begin{center}
{\LARGE\bf Universal description of radially excited heavy and
light vector mesons}
\end{center}

\begin{center}
{\large S. S. Afonin and I. V. Pusenkov}
\end{center}

\begin{center}
{\small V. A. Fock Department of Theoretical Physics,
Saint-Petersburg
State University, 1 ul. Ulyanovskaya, St. Petersburg, 198504, Russia\\
}
\end{center}

\begin{abstract}
A new qualitative stringlike picture for mesons is proposed which
leads to a simple and intuitively clear generalization of linear
radial Regge trajectories to the case of massive quarks. The
obtained universal relation is successfully tested in the sector
of unflavored vector mesons, where many radial excitations are
known. Some new predictions are given. Our results suggest that
the quark masses can be easily estimated from the spectra of
radially excited mesons.
\end{abstract}

\section{Introduction}

Starting from early times of the hadron spectroscopy it has been
widely believed that the dynamics responsible for the formation of
hadrons is more or less universal at all scales where the hadron
resonances are observed. The discovery of QCD provided a powerful
support for this belief. Unfortunately, QCD is still not amenable
to analytical calculation of the hadron spectrum and for this
purpose one resorts to simplified dynamical models simulating QCD.
In practice, for description of the light and heavy hadrons, one
exploits usually different models. The main problem in the hadron
spectroscopy, however, is the building of a universal solvable
model describing in detail the whole meson or baryon spectrum as a
function of quark masses. Despite a great deal of interesting
attempts (see, e.g.,~\cite{rujula,kang,isgur}), such a
quantitative model has not been constructed.

The question, then, is where should we look for the signs of
universality? It is hardly believable that we can observe it in
decay processes (in decay width) which depend on the number of
quark flavors and on the energy scale. But we may hope to find
some interesting manifestations of the universality in the mass
spectrum of hadron states. Our hope is based on the observation of
approximately linear Regge, $M^2\sim J$, and linear radial Regge,
$M^2\sim n$, trajectories in the heavy mesons (see discussions
in~\cite{likhoded}). Here $M$ is the meson mass, $J$ denotes the
spin, and $n$ means the radial quantum number enumerating the
daughter Regge trajectories. The point is that in the sector of
light mesons, such linear trajectories were predicted long ago by
the Veneziano dual models and by the hadron string models. The
experimental verification appeared much later. The first (and
still the only) phenomenological evidence came from the combined
analysis of experimental data on proton-antiproton annihilation
obtained by the Crystal Barrel Collaboration~\cite{ani}. The most
of light nonstrange mesons discovered in this analysis are still
among not confirmed states in the compilation of Particle Data
Group~\cite{pdg}. However, according to the review~\cite{bugg},
the existence of many of the discovered resonances is quite
secure.

The spectrum of light mesons enriched by the numerous states
extracted from the Crystal Barrel data has been extensively used
in many models for fitting the model parameters and for comparison
of theoretical predictions with the real phenomenology. The
deviations from linearity in the light mesons were studied within
the framework of large-$N_c$ QCD sum rules~\cite{sr} and in the
holographic approach~\cite{UVholog}. In the heavy sector, such
deviations were analyzed in Ref.~\cite{likhoded}. Concerning the
recent progress in understanding the global spectral behavior, it
was proposed that the spectrum of light nonstrange mesons
possesses a large degeneracy of the kind $M^2\sim J+n$~\cite{cl1}.
Later it was argued that the real degeneracy has the form $M^2\sim
L+n$, where $L$ is the orbital momentum of the valent
quark-antiquark pair~\cite{clust_rev}. This observation was also
independently made in Refs.~\cite{klempt,shif_cluster}. Physically
the degeneracy in question means the existence of a "principal"
quantum number in light mesons~\cite{parity} which coincides with
that of the hydrogen atom. The corresponding hydrogenlike
classification of meson states was developed in Ref.~\cite{hydr}.
The discussions of some further consequences are given
in~\cite{conseq}. Recently an alternative analysis gave the
averaged spectrum $M^2\sim J+1.23n$~\cite{arriola}. Many
illuminating discussions concerning the analysis~\cite{arriola}
are contained in Refs.~\cite{bugg2012,arriola_b}.

The extension of these discussions to the heavy mesons would be
inevitably very speculative because the variety of established
states in this sector is much more restricted. First of all, the
highest observed spin of heavy mesons is $J=2$ while that of the
light mesons is $J=6$. Second, the radial excitations are usually
poorly known. The only exception represents the case of unflavored
vector states. These resonances have quantum numbers of the photon
and therefore are intensively produced in the
$e^+e^-$-annihilation. Since the mechanism of creation of the
vector mesons in the $e^+e^-$-annihilation is expected to be
identical for any flavor and abundant experimental data are
available, the sector of unflavored vector mesons is an
appropriate place for analyzing the manifestations of universality
of the strong interactions in the resonance formation.

In the present paper, we study the spectroscopic universality
among the vector mesons with hidden flavor --- the $\varphi$,
$\psi$, $\Upsilon$ mesons and their analogues in the sector of
$u,d$ quarks --- $\omega$ mesons. The paper is organized as
follows. The available data on the unflavored vector mesons are
summarized in Section~2. In Section~3, we recall a simple string
description of the light mesons which leads to the linear radial
Regge trajectories. A straightforward inclusion of heavy quarks
into this description results in a strong nonlinearity of
trajectories. We propose an alternative model based on static
quarks. The obtained spectrum is fitted in Section~4 and the
results are discussed. Section~5 concludes our analysis.

\section{The vector spectrum}

The masses of known unflavored vector mesons are given in Table~1.
According to the quark model, the vector two-quark states can
represent the $S$ or $D$-wave resonances. We tried to keep the
typical assignment to the $S$ or $D$ states accepted in the
literature.
\begin{table}
\caption{\small The masses of known $\omega$, $\phi$, $\psi$ and
$\Upsilon$ mesons (in MeV)~\cite{pdg}. The experimental error is
not displayed if it is less than 1~MeV.}
\begin{center}
{
\begin{tabular}{|c|ccccc|}
\hline
$M\setminus n$ & $0$ & $1$ & $2$ & $3$ & $4$\\
\hline
$M_\omega$ \,\,\, $S$& $783$ & $1425(25)$ & --- & --- & $2205(30)$ \\
\,\,\,\,\,\,\,\,\,\,\,\,\, $D$ & $1670(30)$ & $1960(25)$ & $2290(20)$ & --- & --- \\
$M_\phi$ \,\,\, $S$ & $1020$ & $1680(20)$ & --- & $2175(15)$ & --- \\
$M_\psi$ \,\,\, $S$ & $3097$ & $3686$ & $4039(1)$ & $4421(4)$ & --- \\
\,\,\,\,\,\,\,\,\,\,\,\,\, $D$ & $3773$ & $4153(3)$ & --- & --- & --- \\
$M_\Upsilon$ \,\,\, $S$ & $9460$ & $10023$ & $10355$ & $10579(1)$ & $10876(11)$ \\
\hline
\end{tabular}
}
\end{center}
\end{table}

Some doubtful states are omitted. The $\rho(1570)$ is not included
since it seems to be a OZI-violating decay mode of $\rho(1700)$
(see the discussions on this state in Particle Data~\cite{pdg}).
We omit $\omega(2330)$ because it likely represents $\omega(2290)$
obtained by other collaboration (in Table~1, we use Bugg's
analyses of the Crystal Barrel data~\cite{pdg}). We also exclude
$\Upsilon(11020)$ whose assignment is uncertain (its
electromagnetic coupling is suspiciously small in comparison with
the coupling of $\Upsilon(10860)$ thus indicating on a strong
$D$-wave admixture in this resonance).

The $\phi(2175)$ is regarded as the 3rd radial excitation of
$\phi(1020)$ since in this case we have a natural mass splitting
between the 1st and 3rd excitations, about 500~MeV (a natural
splitting between the 1st and 2nd excitations lies near
300--350~MeV). This assignment will be also justified by our fits
in Section~4.

The reliable $D$-wave $\phi$ and $\Upsilon$ are not known. For
this reason we will look for a unified description of the $S$-wave
sector only. The established $S$-wave radial spectra of $\psi$ and
$\Upsilon$ are richer than the spectra of light vector mesons.
They are displayed in Figs.(1a) and (1b).
\begin{figure}[ht]

    \begin{minipage}[ht]{0.46\linewidth}
    \includegraphics[width=1\linewidth]{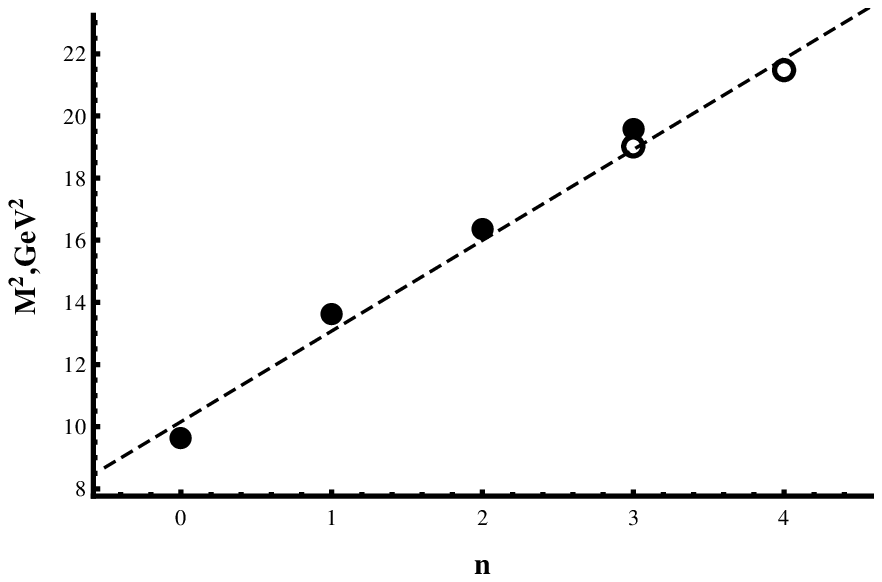} \\
    {\scriptsize Fig. (1a). A presumable spectrum of $S$-wave $\psi$ mesons.
    The experimental points are taken from Table 1. The circles stay
    for $\psi(4361)$ and $\psi(4634)$ observed recently by Belle
    Collaboration~\cite{belle2007,belle2008}.}
    \end{minipage}
    \hfill
    \begin{minipage}[ht]{0.46\linewidth}
    \includegraphics[width=1\linewidth]{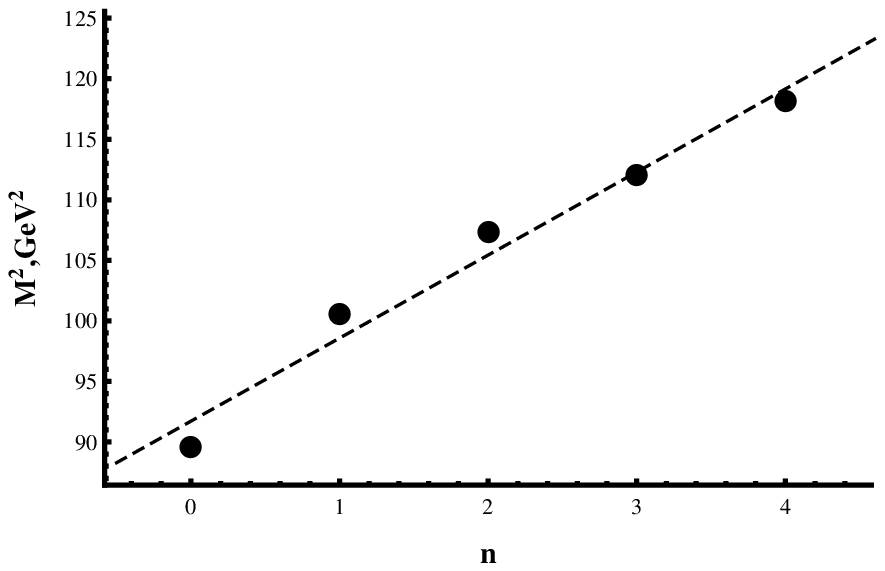} \\
    {\scriptsize Fig. (1b). The spectrum of $S$-wave $\Upsilon$ mesons.}
    \end{minipage}

\end{figure}
These spectra reveal a remarkable Regge behavior
\begin{equation}
\label{1}
M_n^2=a(n+b),
\end{equation}
which is very similar to the one observed in the light nonstrange
mesons~\cite{ani,bugg}. However, the fitted slope $a$ and
intercept $b$ are much larger than those in the light
mesons~\cite{ani,bugg}, see Table~2.
\begin{table}
\caption{\small The radial Regge trajectories~\eqref{1} (in
GeV$^2$) in Figs.(1a) and (1b).}
\begin{center}
{
\begin{tabular}{|c|c|}
\hline
$M_n^2$ & $a(n+b)$ \\
\hline
$M_\psi^2$ & $3.26 (n+3.03)$ \\
$M_\Upsilon^2$ & $6.90 (n+13.28)$ \\
\hline
\end{tabular}}
\end{center}
\end{table}
While for the intercept this looks natural, the slope is
proportional to the tension of hadron string (or energy per unit
length in the potential models with linearly rising confinement
potential) and should have a value of about 1~GeV$^2$.

Below we motivate a linear parametrization for the whole spectrum
of unflavored vector mesons with universal parameters $a$ and $b$.
The only varying parameter will be the quark mass.

\section{Hadron strings}

It is instructive to see how the relation~\eqref{1} emerges in the
stringlike models. There exist plenty of models of this kind in
the literature but some common basic steps can be extracted and
exposed in a simplified manner. One assumes that a massless
quark-antiquark pair (produced, e.g., in the
$e^+e^-$-annihilation) moves back to back creating an elongated
flux-tube of chromoelectric field which one treats as a string.
The energy of the system (the meson mass) is
\begin{equation}
\label{2}
M=2p+\sigma r,
\end{equation}
where $p$ denotes the quark momentum, $r$ is the distance between
the quark and antiquark, and $\sigma$ represents a constant string
tension. One assumes also that the quarks perform the oscillatory
motion inside the flux-tube and that the semiclassical
Bohr-Sommerfeld quantization condition can be applied to this
motion,
\begin{equation}
\label{3}
\int_0^l pdr=\pi(n+b),\qquad n=0,1,2,\dots.
\end{equation}
Here $l$ is the maximal quark-antiquark separation and the
constant $b$ depends on the boundary conditions. Substituting the
momentum $p$ from~\eqref{2} to~\eqref{3} and using the definition
$\sigma=M/l$ one arrives at the expression
\begin{equation}
\label{4}
M_n^2=4\pi\sigma(n+b),
\end{equation}
which has the form of~\eqref{1}.

The massive quarks are introduced via the replacement
\begin{equation}
\label{5}
p\rightarrow\sqrt{p^2+m^2}.
\end{equation}
If we make the substitution~\eqref{5} in the simplistic model above
(for two quarks of equal mass $m$) we obtain
\begin{equation}
\label{6}
M_n\sqrt{M_n^2-4m^2}+4m^2\ln\frac{M_n-\sqrt{M_n^2-4m^2}}{2m}=4\pi\sigma(n+b).
\end{equation}
In the relativistic limit, $M_n\gg m$, the relation~\eqref{6}
reduces to~\eqref{4} while in the nonrelativistic one,
$M_n-2m\ll2m$, the relation~\eqref{6} yields the spectrum of
linearly rising potential, $M_n\sim n^{\frac23}$. In the general
case of polynomially rising potential $V\sim r^{\alpha}$
($\alpha>0$) one has the spectrum $M_n\sim
n^{\frac{2\alpha}{\alpha+2}}$ at large enough $n$~\cite{brau}. The
spectrum of $\psi$ and $\Upsilon$-mesons cannot be well fitted by
the relation~\eqref{6} with universal parameters $\sigma$ and $b$
because of strong nonlinearity at typical masses of heavy quarks.
This nonlinearity originates from the substitution~\eqref{5}. The
use of this substitution is common in the stringlike~\cite{string}
and semirelativistic potential
models~\cite{kang,isgur,potential,entem} as well as in models
based on the Bethe-Salpeter equation~\cite{salpeter}. All these
models share the nonlinearity at large $m$, although the form of
this nonlinearity is model-dependent. In essence, the quarks are
treated as if they were almost free on-shell particles. If we want
to have a Regge-like behavior for any quark mass, the
substitution~\eqref{5} should not be exploited.

In reality, even the light quarks are always massive. Hence, we
can choose a reference frame where one of quarks is at rest.
Consider the large-$N_c$ limit. The hadrons do not decay and
represent bound states. We assume that the meson bound state
corresponds to the situation when two quarks are situated at a
fixed relative distance. They are bound by exchange of some
massless particle. The physical interpretation for this particle
may be different --- it could be the massless pion or gluon. The
relation~\eqref{2} is then replaced by
\begin{equation}
\label{7}
M=m_1+m_2+p+\sigma r.
\end{equation}
Here $p$ is the momentum of the particle mediator, $m_1$ and $m_2$
are quark masses, $m_1=m_2\equiv m$ for the unflavored mesons. Let
us apply the quantization condition~\eqref{3} to the exchanged
particle. Taking into account that now $\sigma=(M-2m)/l$, this
will result in the mass relation
\begin{equation}
\label{8}
(M_n-2m)^2=2\pi\sigma(n+b).
\end{equation}
The slope $2\pi\sigma$ in~\eqref{8} coincides with the slope of
rotating Nambu string~\cite{nambu}. Thus the model is able to
explain the large degeneracy $M^2\sim J+n$ discussed in
Introduction. A more important point for us is that the spectrum
has the Regge-like form~\eqref{1} with universal parameters
$a=2\pi\sigma$ and $b$. The only varying parameter is the quark
mass $m$. In the next Section, we estimate the free parameters
from the experimental data.

\section{Discussions, fits, and predictions}

Our main result, Eq.~\eqref{8}, was deduced in the large-$N_c$
limit of QCD. This relation predicts infinitely many excited
states as expected in the large-$N_c$ world. In the real $N_c=3$
world, the hadron string breaks down at some point (for
sufficiently large radial number); hence, the relation~\eqref{8}
should have a limit of application. The rigorous derivation of
this limit is a difficult task. But we can propose a qualitative
estimate. The onset of the continuum spectrum may be identified
with the point where the total widths of neighboring resonances
overlap almost completely. Let us accept the following criterion:
If since some excitation number $k$, the half-width is comparable
with the mass difference between the $k$th resonance and the
$(k+1)$th one,
\begin{equation}
\label{add1}
\Gamma_k/2\approx\sqrt{a(k+1+b)}-\sqrt{a(k+b)},
\end{equation}
the $(k+1)$th state is indistinguishable from continuum. The total
decay width (inverse life-time) is proportional to the probability
of creation of extra quark-antiquark pair along the string which,
in turn, is proportional to the string length $l$. As long as $l$
is proportional to the meson mass in this scheme, $M_k=\sigma
l_k$, we conclude~\cite{shifman2000}
\begin{equation}
\label{add2}
\Gamma_k=c_k\sqrt{a(k+b)}.
\end{equation}
The averaged empirical value of constants $c_k$ for the excited
light nonstrange mesons is $c\equiv\langle c_k\rangle\approx 0.1$
(see, e.g., the second reference in~\cite{cl1}). Let us
insert~\eqref{add2} into~\eqref{add1} and treat $c$ as a small
parameter. Keeping only linear in $c$ contribution, we obtain the
estimate
\begin{equation}
\label{add3}
k\approx[1/c-b],
\end{equation}
where the square brackets denote the integer part. Since
$c=\mathcal{O}(1/N_c)$, the number of observable resonances is
$\mathcal{O}(N_c)$. For the light vector states, the constant $c$
lies in the interval $c=0.1-0.2$ while $0\lesssim b\lesssim 1$
(see, e.g., the Fit~A in~Table~3). Then the number of radially
excited states described by the relation~\eqref{8} ranges from
$k=4$ (pessimistic estimate) to $k=9$ (optimistic estimate).

After this caveat, we should test the Eq.~\eqref{8} using the
available experimental data. A traditional problem emerging at
this point is the choice of data; in our case this is determining
which states should be treated as reliable $S$-wave vector
resonances. Another source of uncertainty comes from the fact that
formally we should use the values of meson masses in the
large-$N_c$ limit of QCD which are unknown. The hadron masses
presented in the PDG~\cite{pdg} refer to $N_c=3$ world. They have
uncertainty falling into the interval $M\pm \Gamma/2$ since the
resonance mass $M$ depends on the reaction channel and on the
method used to extract it~\cite{arriola}. The lattice calculations
of meson masses at large values of $N_c$ could help but a little
is known on this subject. The lattice simulations of
Ref.~\cite{bali} resulted in the following interpolating formula
for the $\rho$-meson mass in the chiral limit:
$m_\rho/\sqrt{\sigma}=1.54(1)+0.92(21)/N_c^2$, where $\sigma$
represents the universal string tension measured on the lattice.
This relation seems to predict that the $\rho$-meson mass
diminishes by about 50~MeV when going to $N_c=\infty$ world plus
small chiral corrections. The large-$N_c$ studies in the
unitarized chiral perturbation theory~\cite{unitar} led to the
opposite effect --- the enhancement of $m_\rho$ by 40-60~MeV in
the large-$N_c$ limit. But both estimates fall into the
uncertainty interval $m_\rho \pm \Gamma_\rho/2$
($\Gamma_\rho=147.8(9)$~MeV~\cite{pdg}).

Since our analysis is rather qualitative and we do not pretend to
the accuracy better than the accuracy of large-$N_c$ limit in QCD,
it is enough to present one reasonable fit based on the
relation~\eqref{8}. We have tried many possibilities and
assumptions. Below two typical fits are provided. In both cases
the spectrum of $\omega$ mesons is used for the light nonstrange
vector resonances ($\omega$ is the direct analogue of unflavored
quarkonia as the flavor part of its wave function is symmetric).
Within the approximate relation~\eqref{8}, the $\omega$ and $\rho$
are degenerate. Phenomenologically this is consistent with the
accuracy of the large-$N_c$ limit.

In the first case (Fit A), we assume the $M^2\sim L+n$ degeneracy
in the light nonstrange mesons~\cite{clust_rev}. According to the
analysis of the first reference in~\cite{clust_rev}, the
phenomenological spectrum is (in GeV$^2$)
\begin{equation}
\label{9}
M^2(L,n)\approx1.10(L+n+0.62).
\end{equation}
The given degeneracy allows us to replace the masses of unknown
$S$-wave $\omega$ mesons in Table~1 by the masses of corresponding
known $D$-wave states. The masses of $u$ and $d$ quarks are set to
zero (the chiral limit is accepted for simplicity). It is of
course implied that in reality the $u$ and $d$ quarks have small
masses (this was assumed in the derivation of relation~\eqref{8}),
we just neglect the $\mathcal{O}(m_{u,d})$ shifts in the meson
masses which are expected to be much smaller than the errors
induced by the departure from the large-$N_c$ limit of QCD.

The results of global fit are displayed in Table~3. We see a
striking agreement of the obtained parameters both with the
phenomenological spectrum~\eqref{9} and with the current-quark
masses in Particle Data~\cite{pdg}: $m_s=95(5)$~MeV,
$m_c=1.275(25)$~GeV, $m_b=4.18(3)$~GeV. It should be taken into
account that these masses are given at a scale 2~GeV. Since the
typical scale of excited $\phi$ mesons is rather 1~GeV per quark,
another value of $m_s$ should be used for comparison:
$m_s(1~\text{GeV})\approx1.35m_s(2~\text{GeV})\approx128(7)~\text{GeV}$.
It should be noted also that if we ascribe $\phi(2175)$ to the 2nd
radial excitation of $\phi(1020)$, the global fit is worse.
\begin{table}
\caption{\small The quark masses (in GeV), the slope $a$ (in
GeV$^2$) and the dimensionless intercept parameter $b$ in the
relation~\eqref{8}.}
\begin{center}
{\footnotesize
\begin{tabular}{|c|cc|}
\hline
 & Fit A & Fit B\\
\hline
$m_{u,d}$ & 0 & 0.28(4)\\
$m_s$ & 0.12(8) & 0.40(3) \\
$m_c$ & 1.20(7) & 1.48(5)\\
$m_b$ & 4.32(6) & 4.59(5)\\
\hline
$a$ & 1.06(3) & 0.67(7)\\
$b$ & 0.63(7) & 0.00(0)\\
\hline
\end{tabular}}
\end{center}
\end{table}

In principle, the demonstration of this fit is sufficient to
justify the viability of the relation~\eqref{8}. But we would like
to show an alternative interpretation for the parameter $m$
in~\eqref{8}. Let us leave the assumption about large
degeneracy~\eqref{9} in the light mesons and determine the
parameters $a$, $b$, $m_c$, $m_b$ from the sector of heavy
quarkonia where the radial spectrum is better established. After
that we obtain the parameters $m_{u,d}$ and $m_s$ from the best
fit to the known $S$-wave $\omega$ and $\phi$ mesons. The results
are given in Table~3 (Fit B). According to this fit, it looks
natural to interpret the parameter $m$ in~\eqref{8} as the mass of
constituent quark. Indeed, the constituent mass of $u,d$ quarks
lies around 300~MeV, the difference $m_s-m_{u,d}=120$~MeV is close
to the current mass of the strange quark as it is expected in the
potential models\footnote{This is not the case if we ascribe
$\phi(2175)$ to the 2nd radial excitation of $\phi(1020)$. We
would then obtain $m_s=0.49(3)$~GeV. The given observation is
another one justification for our interpretation of
$\phi(2175)$.}, the so-called $1S$-mass of $b$-quark is
4.66(3)~GeV~\cite{pdg}. It should be noted that masses of
constituent quarks can be quite different in different potential
models. For instance, $m_{u,d}=220$~MeV in Ref.~\cite{isgur} and
$m_{u,d}=313$~MeV in~\cite{entem}.

If we relax the condition $m_{u,d}=0$ in the Fit~A and take
$m_{u,d}$ as a free parameter, the ensuing fits resemble the
Fit~B, the parameter $m$ acquires then typical values of
constituent masses in the potential models.

A graphical comparison of the relation~\eqref{8} with the
experimental spectra is plotted in Figs.~(2a) and~(2b). Some meson
masses predicted by the Fits~A and~B are demonstrated in
Tables~4--7. The measure of quality of the fits is
$\chi^2=\sum_n\left(\frac{M_n^2-M_{n,\text{exp}}^2}{M_{n,\text{exp}}^2}\right)^2$.

As we have mentioned, other assumptions for fitting the
experimental data are possible. For example, one can notice that
the ground state in Figs.~(1a) and (1b) lies systematically below
the linear trajectory. This observation may motivate the exclusion
of the ground state in our analysis. However, excluding the ground
state does not lead to a noticeable improvement of the global fits
and predictions. In any case, the relation~\eqref{8} remains
reasonable from the phenomenological point of view, albeit
parametrically dependent on the choice of data.
\begin{figure}[ht]
    \begin{minipage}[ht]{0.46\linewidth}
    \includegraphics[width=1\linewidth]{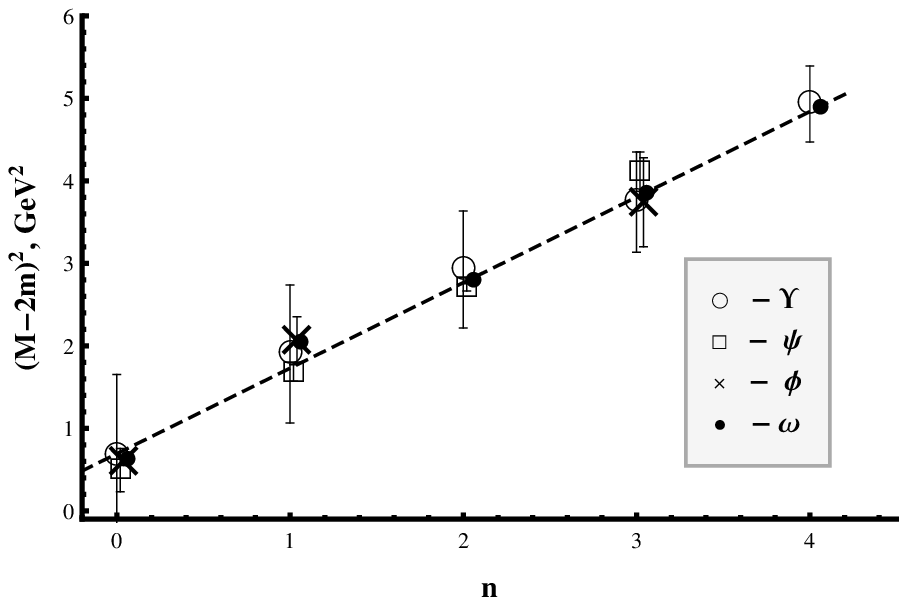} \\{\scriptsize Fig. (2a). The spectrum (8) for the Fit A.}
    \end{minipage}
    \hfill
    \begin{minipage}[ht]{0.46\linewidth}
    \includegraphics[width=1\linewidth]{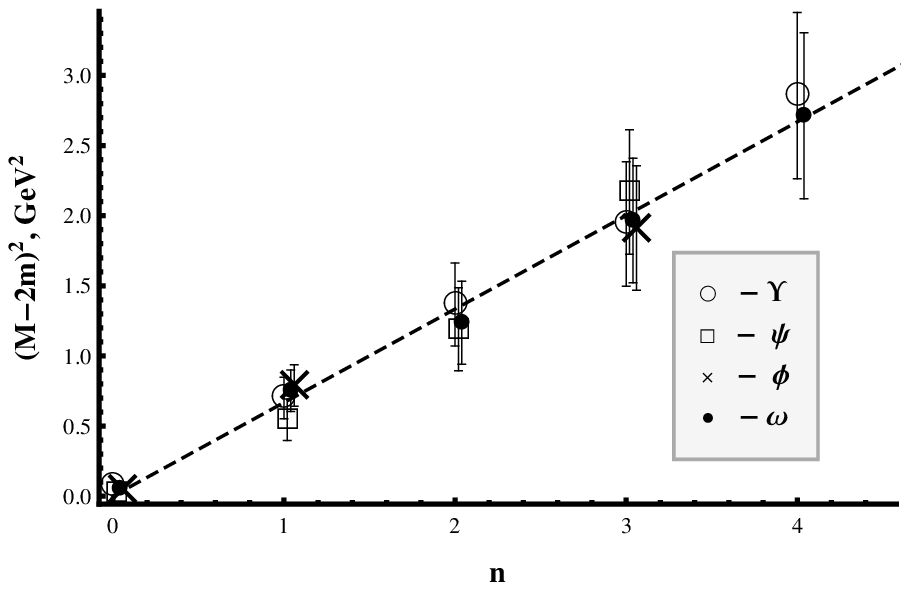} \\{\scriptsize Fig. (2b). The spectrum (8) for the Fit B.}
    \end{minipage}
\end{figure}

\begin{table}
\caption{\small The theoretical and experimental masses of
$S$-wave $\omega$ mesons (in GeV)~\cite{pdg}. The experimental
error is not displayed if it is less than 1~MeV. The question mark
stays at the states whose assignment is questionable.}
\vspace{-0.7cm}
\begin{center}
{
\begin{tabular}{|c|cccccc|c|}
\hline
$M_\omega(n)\setminus n$ & $0$ & $1$ & $2$ & $3$ & $4$ & $5$ & $10^3\chi^2$ \\
\hline
Fit A & $0.82(5)$ & $1.32(3)$  & $1.67(3)$ & $1.96(3)$ & $2.22(3)$ & $2.44(3)$ & $39$ \\
Fit B & $0.56(8)$ & $1.38(9)$ & $1.72(10)$ & $1.98(11)$ & $2.20(12)$ & $2.39(12)$ & $249$ \\
Experiment & $0.783$ & $1.425(25)$ & $1.67(30)^?$ & $1.960(25)^?$ & $2.205(30)$ & $2.330(30)^?$ &\\
\hline
\end{tabular}
}
\end{center}
\end{table}
\begin{table}
\caption{\small The theoretical and experimental masses of
$S$-wave $\phi$ mesons (in GeV)~\cite{pdg}.} \vspace{-0.7cm}
\begin{center}
{
\begin{tabular}{|c|cccccc|c|}
\hline
$M_\phi(n)\setminus n$ & $0$ & $1$ & $2$ & $3$ & $4$ & $5$ & $10^3\chi^2$ \\
\hline
Fit A & $1.06(16)$ & $1.56(16)$ & $1.91(16)$ & $2.20(16)$ & $2.46(16)$ & $2.68(16)$ & $26$ \\
Fit B & $0.80(6)$ & $1.62(7)$ & $1.96(8)$ & $2.22(10)$ & $2.44(10)$ & $2.63(11)$ & $155$ \\
Experiment & $1.020$ & $1.680(20)$ & --- & $2.175(15)$ & --- & --- & \\
\hline
\end{tabular}
}
\end{center}
\end{table}
\begin{table}
\caption{\small The theoretical and experimental masses of
$S$-wave $\psi$ mesons (in GeV)~\cite{pdg}.} \vspace{-0.7cm}
\begin{center}
{
\begin{tabular}{|c|cccccc|c|}
\hline
$M_\psi(n)\setminus n$ & $0$ & $1$ & $2$ & $3$ & $4$ & $5$ & $10^3\chi^2$ \\
\hline
Fit A & $3.21(15)$ & $3.71(14)$ & $4.06(14)$ & $4.36(14)$ & $4.61(14)$ & $4.84(14)$ & $6$ \\
Fit B & $2.96(10)$ & $3.78(11)$ & $4.12(12)$ & $4.38(12)$ & $4.60(13)$ & $4.79(14)$ & $12$ \\
Experiment & $3.097$ & $3.686$ & $4.039(1)$ & $4.361(9)(9)$ & $4.634(8)(8)$ & --- & \\
\hline
\end{tabular}
}
\end{center}
\end{table}
\begin{table}
\caption{\small The theoretical and experimental masses of
$S$-wave $\Upsilon$ mesons (in GeV)~\cite{pdg}.} \vspace{-0.7cm}
\begin{center}
{
\begin{tabular}{|c|cccccc|c|}
\hline
$M_\Upsilon(n)\setminus n$ & $0$ & $1$ & $2$ & $3$ & $4$ & $5$ & $10^3\chi^2$ \\
\hline
Fit A & $9.46(13)$ & $9.96(13)$ & $10.31(13)$ & $10.61(13)$ & $10.86(13)$ & $11.09(13)$ & $0.4$ \\
Fit B & $9.18(10)$ & $10.00(11)$ & $10.34(12)$ & $10.58(12)$ & $10.82(13)$ & $11.01(14)$ & $3.5$ \\
Experiment & $9.460$ & $10.023$ & $10.355(1)$ & $10.579(1)$ & $10.876(11)$ & $11.019(8)$ & \\
\hline
\end{tabular}
}
\end{center}
\end{table}

The previous version of our analysis, Ref.~\cite{univer_afonin},
motivated the authors of Ref.~\cite{univer_arriola} to study the
universality of radial meson trajectories in a different
large-$N_c$ framework. The best fit was obtained after exclusion
of the ground states. The results turned out to be very close to
our Fit~A except the intercept parameter $b$ ($b=0.93(50)$ in
Ref.~\cite{univer_arriola}). Our aim was not to find the best fit
but rather to demonstrate the viability of the phenomenological
relation~\eqref{8} within the accuracy of the large-$N_c$
approximation. The Fit~A meets our goal. Unfortunately, many
details are missing in the short report~\cite{univer_arriola} that
hampers a comparison of our analysis with that of
Ref.~\cite{univer_arriola}.

The mass of the ground state is seriously underestimated in the
Fit~B where this mass is the sum of constituent quark masses. This
makes the Fit~B worse than Fit~A if we use our $\chi^2$ criterion.
However, if the ground states are excluded from $\chi^2$ then
$\chi^2$ is less (by a factor of 3) for the Fit~B in all cases
except the $\psi$ mesons.

Concluding our discussions, we make some remarks on the
predictions displayed in Tables~4--7.  Our fits indicate the
existence of a new $\phi$ meson in the mass interval
1900--2000~MeV. The resonance $\psi(4361)$ observed by Belle
Collaboration~\cite{belle2007} seems to be a better candidate for
the role of the 3rd radial excitation of the $J/\psi$-meson than
the resonance $\psi(4415)$ (although the latter was used as an
input in our fits). It looks natural to assign the next excitation
to $\psi(4634)$ also observed by Belle Collaboration some time
ago~\cite{belle2008} and interpret $\psi(4415)$ as a $D$-wave
state. This assignment agrees with the results of
Ref.~\cite{entem}. The $\Upsilon(11020)$ is compatible with the
role of the 5th excitation of $\Upsilon(9460)$.

\section{Conclusions}

We put forward a new generalization of linear radial Regge
trajectories to the case of massive quarks. Within this
generalization, the form of contribution to the meson masses due
to confinement is universal for any quarkonia. We demonstrated
that the ensuing mass relation~\eqref{8} can be well consistent
with the experimental data on the unflavored vector mesons. In
addition, it leads to some interesting predictions in the meson
spectroscopy. Although our considerations were simple and did not
take into account many effects (such as the relativistic mixing of
$S$ and $D$ wave states resulting in some mass
shifts~\cite{eichten} and the mass shifts caused by the transition
from the large-$N_c$ limit to the real $N_c=3$ world) the quality
of final fits is comparable with typical results of
semirelativistic potential models and other technically nontrivial
approaches.

A natural question emerging after our analysis is whether the
relation~\eqref{8} can be extended to other types of mesons?
Unfortunately, the available experimental data are too scarce for
making any definite conclusions. Among the possible extensions
of~\eqref{8} are
\begin{equation}
\label{10}
(M_n-m_1-m_2)^2=a(n+x+b),
\end{equation}
where we may have $x=\alpha L$ or $x=\alpha J$ (here $L$ and $J$
mean the orbital momentum of valent quarks and total spin). The
constant $\alpha$ should be fixed from the phenomenology. A
intriguing possibility $\alpha=1$ would lead to a large degeneracy
observed in the light nonstrange mesons. Needless to say,
establishing a consistent form for the relation~\eqref{10} is very
interesting since it is able to give many predictions for new
experiments in the hadron spectroscopy. This problem is left for
future work.

\section*{Acknowledgments}

The authors acknowledge Saint-Petersburg State University for
Research Grant No. 11.38.189.2014. The work was also partially
supported by the RFBR Grant No. 13-02-00127-a.

\end{document}